\documentclass[12pt]{iopart}

\usepackage{graphicx}
\usepackage{wrapfig}
\usepackage{sidecap}
\begin{document}

\title[Diffraction dissociation in $pp$ collisions with ALICE]{
Diffraction dissociation in proton-proton collisions at $\sqrt{s}$ = 
0.9 TeV, 2.76 TeV and 7 TeV with ALICE at the LHC}
\author{M.G. Poghosyan for the ALICE collaboration}
\address{
Universit\'a  di Torino/INFN, 10125 Torino, Italy}
\ead{martin.poghosyan@cern.ch}
\begin{abstract}
The relative rates of single- and double- diffractive processes were measured 
with the ALICE detector by studying properties of gaps in the pseudorapidity 
distribution of particles produced in proton-proton collisions at $\sqrt{s}$
 = 0.9 TeV, 2.76 TeV and 7 TeV. ALICE triggering efficiencies are determined 
for various classes of events, using a detector simulation validated with data 
on inclusive particle production. Cross-sections are determined using van der 
Meer scans to measure beam properties and obtain a measurement of the luminosity.
\end{abstract}
\vspace*{-1.cm}
\section{Introduction}
\vspace*{-0.5cm}
The phenomenon of diffractive dissociation was predicted by E.L. Feinberg and I.Ya. Pomeranchuk 
\cite{FeinbergPomeranchuk} before it was observed experimentally. This process became a subject of 
intensive experimental studies at all hadron accelerators including the high energy facilities at CERN and FNAL
(ISR, S$p\bar{p}$S and  Tevatron, and now LHC).\\
Being diffractively produced, the system must have the same intrinsic 
quantum numbers as the incoming hadron while spin and parity may be different because some orbital 
angular momentum can be transferred to the system during the interaction.
Regge theory is the main framework for describing such processes. The diffraction dissociation
process is described by the phenomenology of Pomeron exchange, where the Pomeron is a color 
singlet with quantum numbers of the vacuum.\\
Experimentally, it is not possible to select from large rapidity processes those that are caused 
by a Pomeron exchange. Therefore, we associate the diffraction dissociation with large rapidity 
gap processes, considering the contribution of secondary-Reggeons as well. The separation of these 
processes is model dependent.\\
In experiments (such as ALICE at LHC) where the non-diffracted proton in single-diffraction (SD) is 
outside detector acceptance, the reconstruction of the characteristics of this process becomes 
model dependent. 
Therefore, the physical model which is chosen as an input for data analysis and correction, 
should be as close to 
reality as possible. By reality we mean the available data on total, elastic and diffractive 
interaction 
cross-sections of $pp$ and $p\bar{p}$ collisions provided by experiments performed up to now.\\
A model based on Gribov's Regge calculus was developed \cite{MyDiff} and was proposed to describe 
diffractive processes. The numerical evaluation of the model 
gave a good description of data on diffraction dissociation processes 
in $pp$ and $p\bar{p}$ interactions over a wide energy range (from $P_{lab} = 65$ GeV/$c$ to 
$\sqrt{s} = 1800$ GeV) explored with various accelerators at CERN and at Fermilab \cite{MyDiff}. 
In the measurement described here the model \cite{MyDiff} is used to provide the dependence of 
SD cross-section on diffracted mass in PYTHIA6 \cite{PYTHIA6} and PHOJET \cite{PHOJET} 
Monte Carlo (MC) generators.\\
\vspace*{-1.cm}
\section{Analysis method}
\vspace*{-0.5cm}
A detailed description of the ALICE detector can be found in \cite{ALICE_JINST}. 
In this study we used three of its sub detectors: The Silicon 
Pixel Detector (SPD), the VZERO scintillator modules and the Forward Multiplicity Detector (FMD). 
SPD and VZERO are the main ALICE triggers for collecting minimum bias events. The 
FMD extends the pseudorapidity coverage to the interval from -3.7 to 5.1.\\
We studied, on an event per event basis, the pseudorapidity distribution of tracks made of the event vertex
and a hit in either SPD, VZERO or FMD cells.
For each event we found the pseudorapidity gap with the largest width
  and calculated the pseudorapidity distances $d_1$ ($\eta<0$ side), $d_2$ ($\eta>0$ side) of each edge 
of the measured pseudorapidity 
        distribution from the corresponding nearest edge of the detector acceptance. 
After finding the widest pseudorapidity gap, we classified the events into 
1-arm and 2-arm triggers as follows:
\vspace*{-.3cm}
\begin{itemize}
  \item If the maximum gap width is greater than both $d_1$ and $d_2$,
            the event is classified as 2-arm trigger event. 
\vspace*{-.2cm}
  \item If the edge is at $\eta > -1$ or $\eta < 1$ and $d_1$ or $d_2$  is bigger than the maximum gap width, 
         the event is classified as 
        Left-side or Right-side 1-arm trigger event, respectively. 
\vspace*{-.2cm}
  \item The rest of the events we considered as 2-arm trigger events.  
\end{itemize}
\vspace*{-.2cm}
The fraction of SD processes was measured by counting the relative rate 
of one-arm 
and two-arm triggers.
MC simulations showed that masses above 200~GeV$/c^2$ mainly give 2-arm trigger and therefore for our measurement 
the $M$ =~200~GeV$/c^2$ 
serves as the boundary between the SD and non-single diffractive (NSD) events.\\
Several tests were made to be sure that the material budget and the inefficiency of detectors do not spoil the 
pseudorapidity gaps. 
In particular, we varied the fraction of SD
in MCs and studied the dependence of the measured fraction of SD vs the input fraction of 
SD. We found that there is one to one relation between input and output fractions and 
the cases with real and ideal detectors are very close to each other. We also varied the cross-section of 
double-diffraction (DD) in MCs to study the sensitivity of the pseudorapidity gap 
width distribution in 2-arm trigger events on the input fraction of DD. 
For this case again one to one relation was found.\\
A comparison with data showed that with the default DD fraction PYTHIA significantly overestimates 
the fraction of large pseudorapidity gaps and PHOJET significantly underestimates it.
%
%
In order to have a constraint on the contribution of large rapidity gap NSD events in the one-arm triggers,
the DD fraction 
in PYTHIA and PHOJET was varied. In PYTHIA/PHOJET the default DD fraction  
 is 0.12/0.06  at 900 GeV and 0.13/0.05 at 7 TeV and
 we set it to 0.1/0.11 at 900 GeV and 0.09/0.07 at 7 TeV. 
For the $\sqrt{s} = $ 2.76 TeV run, taken recently, the performance of FMD is not well understood yet. Therefore, 
we used only the
SPD and VZERO detectors.\\ 
\vspace*{-1.cm}
\section{Results}
\begin{figure}[h!]
\begin{minipage}[t]{.5\textwidth}
\begin{center}
\includegraphics[width=.95\textwidth]{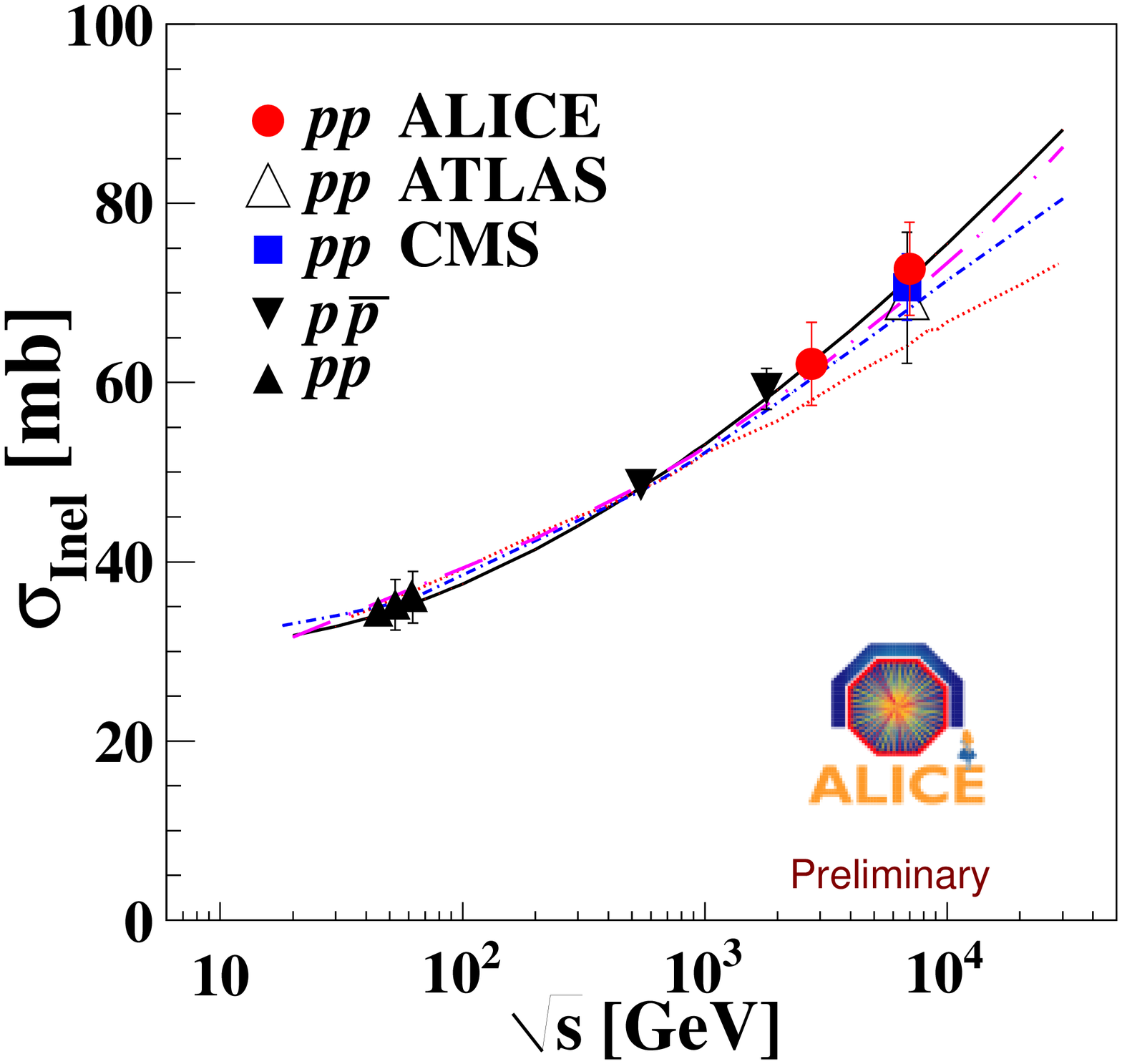}
\end{center}
\hspace{-2.cm}
\begin{minipage}[c]{1.2\textwidth}
\vspace*{-0.5cm}
\begin{center}
\caption{
\label{Fig:XS_Inel}
Inelastic cross-section as a function of collision energy.
Data are compared with the predictions of \cite{MyDiff} (solid black line), 
\cite{Ostapchenko} (long dot-dashed pink line),
\cite{Gotsman} (short dot-dashed blue line)
and \cite{Ryskin} (dotted red line).
Data from other experiments are taken from \cite{InelXSdata}
}
\end{center}
\end{minipage}
\end{minipage}
\hfill
\begin{minipage}[t]{.5\textwidth}
\vspace*{-5.5cm}
\begin{center}
\includegraphics[width=.95\textwidth]{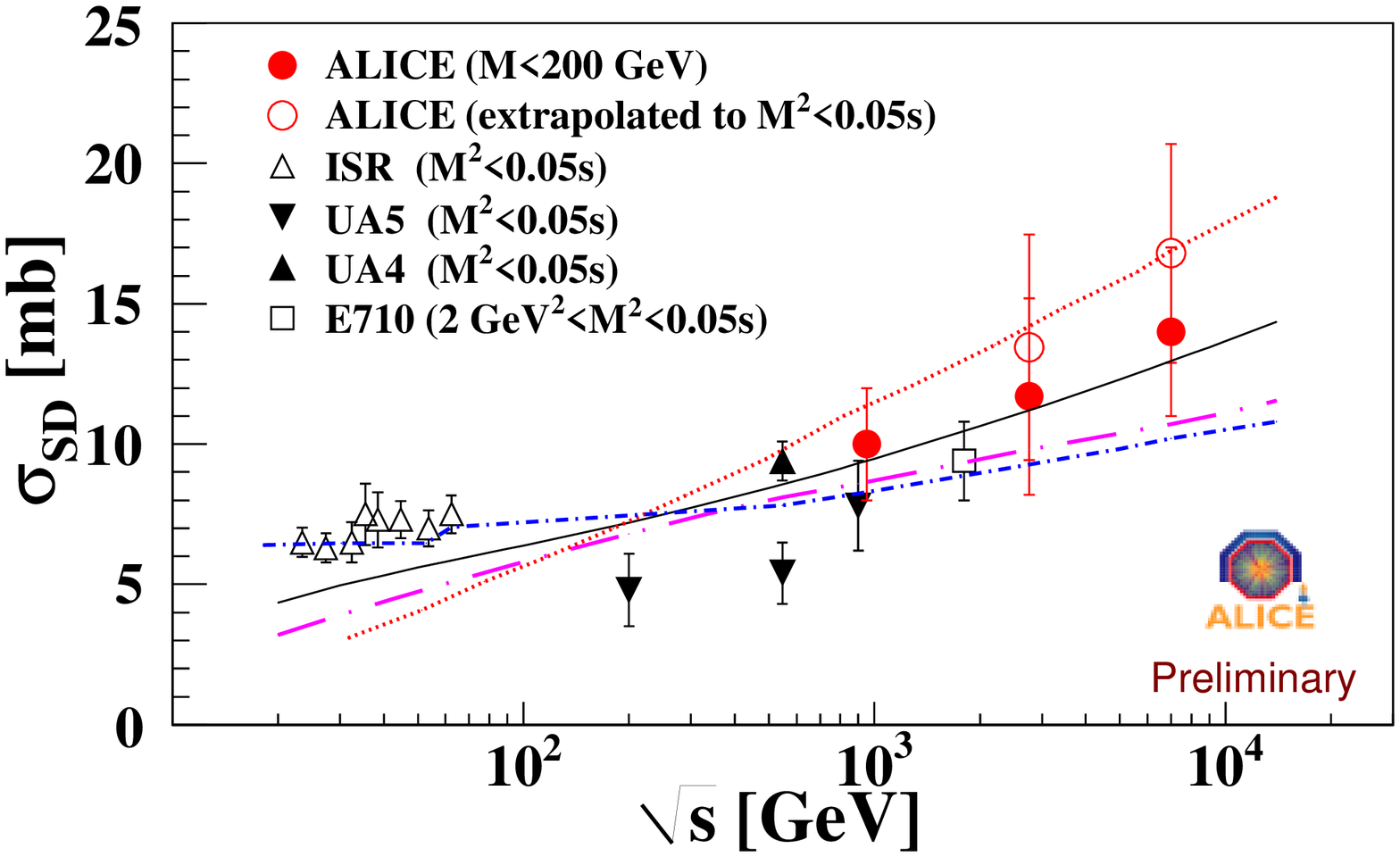}
\end{center}
\hspace{-2.cm}
\begin{minipage}[c]{1.2\textwidth}
\begin{center}
\vspace*{-0.5cm}
\caption{
\label{Fig:XS_SD}
Single-diffractive cross-section as a function of collision energy.
Data from other experiments are for $M^2 < 0.05s$ \cite{SDXSdata}.
ALICE measured points are shown with full (red) circles, and in order to compare  
with data from other experiments were extrapolated to $M^2 < 0.05s$.
The predictions of theoretical models correspond to $M^2 < 0.05s$ and
are defined as in Figure~\ref{Fig:XS_Inel}. 
}
\end{center}
\end{minipage}
\end{minipage}
\end{figure}

\begin{center}
\begin{SCfigure}[][h]
\begin{minipage}[t]{.5\textwidth}
\centering
\begin{minipage}[t]{.5\textwidth}
\includegraphics[width=2.\textwidth]{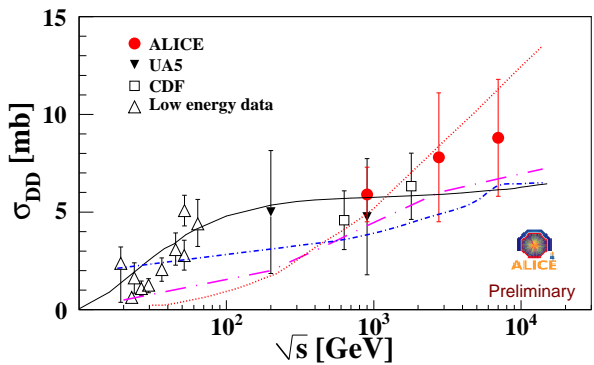}
\end{minipage}
\vspace*{-1.cm}
\caption{
\label{Fig:XS_DD}
Double-diffractive cross-section as a function of collision energy.
The theoretical model predictions are for $\Delta\eta >3$ 
and
are defined as in Figure~\ref{Fig:XS_Inel}.
Data from other experiments are taken from \cite{DDXSdata}.}
\end{minipage}
\end{SCfigure}
\end{center}

\vspace*{-.5cm}
In Table \ref{Tab:tab1} we present the corrected
\begin{table}[h]
\vspace*{-.7cm}
\caption{Fractions of SD ($M <$ 200 GeV/$c^2$) and DD ($\Delta\eta >3$) events.}
\vspace*{-.2cm}
\begin{center}
\begin{tabular}{|c|c|c|c|c|}
\hline
 $\sqrt{s}$ (TeV) & $\sigma_{SD}^{right}/\sigma_{Inel}$ & $\sigma_{SD}^{left}/\sigma_{Inel}$ & $\sigma_{SD}/\sigma_{Inel}$ & $\sigma_{DD}/\sigma_{Inel}$\\
\hline
0.9               &   0.100 $\pm$ 0.015      & 0.102 $\pm$ 0.019   & 0.202 $\pm$ 0.034 &   0.113 $\pm$ 0.029\\
2.76              &   0.090 $\pm$ 0.028      & 0.097 $\pm$ 0.026   & 0.187 $\pm$ 0.054 &   0.125 $\pm$ 0.052\\
7                 &   0.100 $\pm$ 0.020      & 0.101 $\pm$ 0.019   & 0.201 $\pm$ 0.039 &   0.122 $\pm$ 0.036\\
\hline
\end{tabular}
\end{center}
\label{Tab:tab1}
\vspace*{-.8cm}
\end{table}
ratios of single-diffraction over inelastic cross-sections. 
Statistical errors are negligible and the quoted errors are systematic.~They come from the adjustment 
of DD in PYTHIA and PHOJET, 
from changing the $\sigma^{-1} d\sigma/dM$ by $\pm 50 \% $ at the proton-pion mass threshold, 
from the uncertainty of the SD kinematic in PYTHIA and PHOJET and from the beam-gas background.
Despite different acceptances and different trigger ratios of the two ALICE sides, the corrected ratios 
of each side are the same as expected from the symmetry of the process.\\
After tuning MC generators for large rapidity gaps, we calculated the fraction of NSD events
with pseudorapidity gap $\Delta\eta > 3$ (see Table \ref{Tab:tab1}). 
Using the obtained fractions of  SD and DD, we calculated the 
efficiencies of detecting $pp$ inelastic interactions by requiring a coincidence between 
the two sides of the VZERO detectors ($MB_{AND}$)  
and a logical OR between the signals from the SPD and VZERO detectors.
Their ratio was compared with data and a good agreement was found.
For $MB_{AND}$ we obtained 
$(76.2 \pm 2)\%$ and $(74.5 \pm 1.1)\%$ at 2.76 TeV and 7 TeV, respectivelly. 
Using the van der Meer scans to measure the visible cross-section of the $MB_{AND}$ trigger \cite{Ken}, 
and our simulation result for the detector acceptance, for inelastic cross-section we obtained:
$\sigma_{Inel}(2.76 \, TeV) = 62.1 \pm 1.6(model) \pm 4.3(luminosity)$~mb and
$\sigma_{Inel}(7\, TeV)     = 72.7 \pm 1.1(model) \pm 5.1(luminosity)$~mb.
$pp$ inelastic, SD and  DD cross-sections 
are compared with data from other experiments and with the predictions of 
theoretical models from \cite{MyDiff} and \cite{Ostapchenko}-\cite{Ryskin} in Figures \ref{Fig:XS_Inel} - \ref{Fig:XS_DD}.
There is a good agreement between ALICE and UA5 for SD and DD ratios at 900 GeV and between ALICE, ATLAS and CMS for inelastic cross-section at 7 TeV. 
We would like to stress again that in our measurement the diffractive processes are 
associated with large (pseudo)rapidity gap processes. In some measurements and most theoretical models  
the diffraction is considered as a Pomeron exchange, excluding the contribution of Reggeons. \\
\vspace{-1.cm}
\section{Conclusion}
\vspace{-0.5cm}
Fractions of SD ($M < 200$ GeV/$c^2$) and DD ($\Delta\eta > 3$) dissociation processes 
are measured at $\sqrt{s}$ = 0.9, 2.76 and 7 TeV. For $\sqrt{s} = $ 900 GeV a good agreement with UA5 is found. Within our accuracy, we do not observe 
variations of the SD fraction with energy ($\sigma_{SD}/\sigma_{Inel} \simeq 0.2$).\\
$pp$ inelastic cross-section is measured at $\sqrt{s}$ = 2.76 and 7 TeV. The result for 7 TeV is in a good agreement with ATLAS and CMS results.\\


\vspace{-1.5cm}
\section*{References}
\vspace{-.5cm}
\begin{thebibliography}{15}
\bibitem{FeinbergPomeranchuk} E.L. Feinberg and I.Ya. Pomeranchuk, 
 Nuovo Cimento Suppl. {\bf 3} (1956) 652.
\bibitem{MyDiff}
A.B. Kaidalov and M.G. Poghosyan, ArXiv:0909.5156 [hep-ph], Eur. Phys. J. {\bf C67} (2010) 397.
\bibitem{PYTHIA6}
In this analysis Perugia-0 (320) tune is used: P.Z. Skands, 
arXiv:0905.3418[hep-ph].
\bibitem{PHOJET} R. Engel, J. Ranft, S. Roesler, Phys. Rev. D 52, 1459 (1995)
\bibitem{ALICE_JINST} K. Aamodt et al., ALICE Collaboration, JINST {\bf 3} (2008) S08002.
\bibitem{Ken} K. Oyama for the ALICE Collaboration, these proceedings.
\bibitem{Ostapchenko} S. Ostapchenko, arXiv:1010.1869, PR {\bf D83} (2011) 114018.
\bibitem{Gotsman} E. Gotsman et al., arXiv:1010.5323, EPJ. {\bf C74} (2011) 1553.
\bibitem{Ryskin} M.G. Ryskin et al., EPJ. {\bf C60} (2009) 249; {\bf C71} (2011) 1617.
\bibitem{InelXSdata}
L. Baksay et al., Nucl. Phys.{\bf B141} (1978) 1. 
N. A. Amos et al., Nucl. Phys {\bf B262} (1985) 689.
M. Bozzo et al., Phys. Lett. {\bf B147} (1984) 392.
S. Klimenko et al., Report No. FERMILAB-FN-0741, 2003.
G. Aad et al., ArXiv:1104.0326 [hep-ex].
M. Marone, talk at the DIS2011 Workshop, Newport News, VA USA, April 11-15, 2011.
\bibitem{SDXSdata}
J. Armitage et al., Nucl. Phys. {\bf B194} (1982) 365.
D. Bernard et al., Phys. Lett. {\bf B186} (1987) 227.
G.J. Alner et al., Phys. Rep. {\bf 154} (1987) 247.
N.A. Amos et al., Phys. Lett. {\bf B301} (1993) 313.
\bibitem{DDXSdata}
G. Alberi and G. Goggi, Phys. Rep. {\bf 74} (1981) 1.
A. Affolder et al. Phys. Rev. Lett. {\bf 87} (2001) 141802.
\end{thebibliography}
\end{document}